\documentclass[11pt,a4paper]{article}
\usepackage{times}
\usepackage{tgtermes}
\usepackage[T1]{fontenc}
\usepackage[utf8]{inputenc}
\usepackage[french]{babel}
\usepackage{fancyhdr}
\usepackage{setspace}
\usepackage{comment}
\usepackage{algorithm}
\usepackage{algpseudocode}
\usepackage{graphicx}
\usepackage{lipsum}
\usepackage{array}
\usepackage{caption}
\usepackage{multicol}
\usepackage{hyperref}
\usepackage{afterpage}
\usepackage{setspace}
\usepackage{pgffor}
\usepackage{amsmath}
\usepackage{amsmath,amssymb,bm}
\usepackage{graphicx}
\usepackage{tikz,graphics,color,fullpage,float,epsf,caption,subcaption}

\usepackage{titlesec}
\titleformat{\section}
  {\normalfont\fontsize{16}{20}\bfseries}{\thesection}{1em}{}
\titleformat{\subsection}
  {\normalfont\fontsize{16}{20}\bfseries}{\thesubsection}{1em}{}
\titlespacing*{\section}
{0pt}{2.ex plus 1ex minus .2ex}{.0ex}
\titlespacing*{\subsection}
{0pt}{0.ex}{.0ex}

\begin{document}

\begin{center}
\begin{spacing}{2.05}
{\fontsize{18}{18} \selectfont
\bf
Dynamic brittle fracture using Lip-field approach in an explicit dynamics context}
\end{spacing}
\end{center}
\vspace{-1.25cm}
\begin{center}
{\fontsize{12}{14} \selectfont
\bf
Rajasekar Gopalsamy\textsuperscript{a}, Nicolas Chevaugeon\textsuperscript{a} \\
\bigskip
}
\end{center}
{\fontsize{10}{10} \selectfont
a. Ecole Centrale de Nantes, GeM Institute, UMR CNRS 6183,1 rue de la Noe,
44321 Nantes, France
}

\vspace{10pt}

{\fontsize{16}{20}
\bf
Abstract :
}
\bigskip

\textit{This paper aims to investigate the dynamic response of a material body undergoing fracture subjected to high strain rate loading conditions such as impact or explosion. In particular, our focus is limited to softening elastic damage models using Lip-field regularization. The Lip-field approach is a new technique to introduce length scale into the softening damage models. It was first presented for the quasi-static case in \cite{Moes LF, Chevaugeon} and then extended to a one-dimensional dynamic case in \cite{Moes LF dynamics}. This paper extends the application of the Lip-field approach to dynamic fracture in two-dimensional cases. Lip-field approach is a variational approach, in which the potential to be minimized is a non-regularized one. A potential without regularization can result in spurious strain localization, but the Lip-field approach resolves this issue by imposing Lipschitz constraints on the damage field. Our focus is limited to utilizing an explicit staggered scheme to determine the displacement and damage fields. Numerical studies are conducted for two-dimensional cases to assess the dynamic behavior of the proposed model.}

\vspace{28pt}
{\fontsize{14}{20}
\bf
Key words :}  
{\fontsize{14}{20}damage, Lip-field approach,   dynamic fracture, variational approach
}

\section{Introduction}
\medskip
This paper focuses on the fracture response of a brittle material subjected to dynamic loading. Dynamic loadings are experienced during impacts such as crashes or explosions and are short-term loadings. Complex fracture phenomena, including crack initiation, multiple crack branching, crack coalescence, and crack arrest, govern the dynamic failure mechanisms in brittle solids. Unlike in static fracture, the crack propagation speed in dynamic fracture is usually close to the speed of sound in the material. As a result, the dynamic fracture is often associated with rapid energy release, significant damage, and catastrophic failure of the material. Engineering safety analyses greatly benefit from the accurate predictive numerical simulation of complex crack patterns that occur under dynamic impact loading. Explicit time integration schemes provide efficient numerical solutions for such phenomena from a numerical standpoint. This paper aims to outline a variational approach for dynamic fracture using Lip-field regularization. It extends the recent work on the Lip-field approach for quasi-static case \cite{Moes LF, Chevaugeon} and 1D dynamic fragmentation \cite{Moes LF dynamics} to the case of the 2D dynamic setting.

In accordance with the classical methods for quasi-static and dynamic fracture, as explained in \cite{Dugdale, Freund}, the propagation of cracks occurs when the energy release rate reaches a critical value. Defects of the classical Griffith model such as the inability to initiate cracks and the inability to predict crack branching can be overcome in 2 common ways: the cohesive zone model  (CZM) and the diffuse damage model.

In CZM \cite{Dugdale1, Barenblatt}, the fracture process occurs not just at the crack tip but also in an extended region surrounding the crack tip called the cohesive process zone. The macro crack is represented by displacement jumps and the behavior of the cohesive zone is modeled by a traction separation law. In numerical Simulations, the cohesive zone's behavior is typically simulated by incorporating cohesive elements, which are commonly positioned at the interface between finite elements. However, this results in the crack path being limited to the element edges  causing these models to be dependent on mesh orientation if adaptive re-meshing is not employed \cite{Zhou}. Some literature on dynamic fracture simulation using CZM can be found in \cite{Xu, Camacho, Pandolfi}. For e.g., in \cite{Xu}, cohesive zones are introduced between the finite elements throughout the bulk of the material to allow for different possible crack paths. The latter was able to capture the crack branching phenomenon under impact loadings without any ad-hoc criterion. To alleviate mesh dependency, eX-Finite Element Method (X-FEM) \cite{Moes} has been used in \cite{Belytschko} to introduce cohesive elements inside the finite elements for the simulation of dynamic brittle fracture. A comparison of X-FEM and cohesive element method to discretize the cohesive zone, for the case of dynamic brittle fracture, is performed in \cite{Song} and both these methods show similar crack speeds and crack paths for certain cases.

On the other hand, diffuse damage models \cite{Kachanov, Lemaitre} involve the use of an internal damage variable that affects the stress-strain relation. These models in the local form are known to suffer from pathological mesh-sensitive results. Hence non-locality (or regularization) has to be introduced to yield mesh-independent results. There exist several ways in the literature to regularize the local models: for e.g., non-local integral damage models \cite{Lorentz, Pijaudier}, gradient damage model \cite{Nguyen1, Peerlings1} and Thick Level Set (TLS) approach \cite{Moes TLS}. Additionally, the fracture has been reformulated as a problem of minimizing regularized energy in the literature \cite{Francfort1, Francfort2}, leading to the emergence of the variational approach to fracture \cite{Francfort3}. Another class of regularized diffuse damage models that recast the problem into the variational framework is the phase-field approach \cite{Miehe, Ambati} to fracture. All these models introduce length scale into the model to enforce non-locality. They also possess the advantage that they are able to deal with crack nucleation, crack branching, and coalescence without any ad-hoc criterion. Some works on fracture under dynamic loading using phase-field regularization can be found in \cite{Borden, Hofacker, Marigo} and using TLS regularization in \cite{Moreau}.

This paper focuses on the Lip-field approach to fracture under dynamic loading. The Lip-field approach is yet another method to introduce non-locality that has been recently introduced in \cite{Moes LF, Chevaugeon} for the quasi-static case and extended to 1D dynamic fragmentation case in \cite{Moes LF dynamics}. In \cite{Moes LF dynamics}, the Lip-field approach is validated for the 1D dynamic case by comparing the results to an equivalent cohesive zone model. It was shown that the dissipated energies and fragment size for both models follow similar trends.  A comparison of a particular type of phase-field and a Lip-field approach is provided in \cite{Rajasekar} and by deriving some equivalence between both approaches, they were able to yield similar results.  The Lip-field approach is similar to the phase-field approach in the sense that they are both variational in nature. However, in the Lip-field approach, the energy functional to be minimized is a local one (non-regularized) while a non-local (regularized) energy functional is considered for the phase-field approach \cite{Miehe, Ambati}. The regularization in the Lip-field approach is then through the use of Lipschtiz constraints that the damage gradient should be bounded by a critical value. The bounds estimate proved in \cite{Moes LF} provides the portion of the domain over which Lipschitz constraints (non-locality) have to be enforced. The problem remains purely local in the remaining portions of the domain. This greatly reduces the computational time.  Moreover, the energy functional is convex separately with respect to displacement and damage variable and the associated constraints (damage irreversibility and Lipschitz constraints) are also convex. Hence as in the phase-field approach, a staggered scheme is used, resulting in a series of convex optimization problems. 

This paper is organized as follows: The next section describes the classical damage mechanics formulation for the dynamic case in non-regularized form, followed by the description of the Lip-field regularization in Section \ref{sec:lf}. Section \ref{ls} elaborates an efficient way to compute Lipschitz regularized damage field in a continuum setting. In Section \ref{nr}, the details of the explicit time discretization and space discretization along with the algorithm for computing displacement and damage fields are provided. Section \ref{sr} studies the ability of the Lip-field approach to simulate dynamic fracture in 2D cases by means of numerical examples. Finally, Section \ref{conc} concludes the paper.

\section{The mechanical model: Lip-field regularization}
 \medskip

 In this section, the time-discretized incremental potential associated with the problem of damage in elasticity under dynamic loading is derived. It will be then seen that the minimization of this incremental potential gives the solution to the mechanical problem.

 We consider the deformation of a material body with homogeneous density $\rho$ and initially occupying a domain $\Omega \in \mathbb{R}^2$. We are interested in finding the deformation $\mathbf{u}$ of the body during the time interval $t \in [0,T] \in \mathbb{R}^+$. Assuming small strain conditions, the small strain $\bm{\varepsilon}$ can be expressed as follows:
 \begin{equation}
     \bm{\varepsilon(u)} = \nabla_s \mathbf{u} = \frac{1}{2} (\nabla \mathbf{u} + \nabla^T \mathbf{u}) \label{eq:eq1}
\end{equation}
where $\nabla$ and $\nabla_s$ are the gradient and symmetric gradient operators.

The initial conditions are  imposed through initial displacement $\mathbf{u_0}$ and  initial velocity $\mathbf{\dot{{u}}_0}$. Regarding the boundary conditions, displacement $\mathbf{u_d}(t)$ is applied to a part of the boundary $\Gamma_u$ and traction forces $\mathbf{t_N}$ applied on the rest of the boundary $\Gamma_N$, such that $\Gamma_u \cup \Gamma_N = \partial \Omega$ and  $\Gamma_u \cap \Gamma_N = \varnothing$. For the displacement field to be kinematically admissible at a given time $t$, it must belong to the space $\mathcal{U}(t)$.
\begin{align}
\mathcal{U}(t) = \{\mathbf{u} \in H^1(\Omega) : \mathbf{u} = \mathbf{u_d}(t)\;\; on \;\;\Gamma_u\}
\label{eq:eq2}
\end{align}
 In the absence of body forces, the equilibrium condition reads as follows.
\begin{align}
    \int_{\Omega} \rho \ddot{\mathbf{u}} \mathbf{u^*} d\Omega\; +\; \int_{\Omega}  \bm{\sigma} :  \bm{\varepsilon}(\mathbf{u^*}) \; d\Omega = \int_{\Gamma_N} \mathbf{t_N} . \mathbf{u^*} dS , \;\;\; \forall \; \mathbf{u^*}\in \mathcal{U}^*
    \label{eq:eq3}
\end{align}
where $\bm{\sigma}$ is the Cauchy stress and $\mathcal{U}^*$ denotes the test function space defined as follows
\begin{align}
    \mathcal{U}^* =\{\mathbf{u} \in H^1(\Omega) : \mathbf{u} = 0\;\; on \;\;\Gamma_u\}
    \label{eq:eq4}
\end{align}
The equilibrium equation given by Eq. (\ref{eq:eq3}) is accompanied by the following initial conditions
\newenvironment{rcases}
  {\left.\begin{aligned}}
  {\end{aligned}\right\rbrace}
\begin{equation}
\begin{rcases}
  \mathbf{u} &= \mathbf{u_0} \hspace{.5cm}\\
  \mathbf{\dot{u}} &= \mathbf{\dot{u}_0} \hspace{.5cm}
\end{rcases}
\hspace{.5cm} in  \;\; \Omega \;\; at \;\; t=0
\label{eq:eq5}
\end{equation}
and the following boundary conditions
\begin{equation}
\begin{rcases}
 \mathbf{u} &= \mathbf{u_d}(t) \; & on \; &\Gamma_u \hspace{.5cm}\\
  \bm{\sigma}.\mathbf{n} &= \mathbf{t_N}(t)  \; & on \; &\Gamma_N \hspace{.5cm}
\end{rcases}
\hspace{.5cm}  at \;\; \; t \in [0,T]
\label{eq:eq6}
\end{equation}
where $\mathbf{n}$ is the outward unit normal vector to the boundary $\partial \Omega$ of the domain $\Omega$.

To complete the set of equations above, it is necessary to supplement the equilibrium equation with the constitutive model. For this purpose, we utilize the approach of generalized standard materials, as outlined in \cite{Germain, Halphen}. The state variables are $\{ \bm{\varepsilon}, d\}$, where $d$ is the internal scalar damage variable taking a value between 0 and 1. The constitutive model is then characterized by two convex thermodynamic potentials: a free energy potential $\psi(\bm{\varepsilon},d)$ and a dissipation potential $\phi(d,\dot{d})$. For the considered brittle material, the two potentials are defined as follows:
\begin{align}
    \psi &=  g(d)   \left( \frac{\lambda}{2}  <tr(\bm{\varepsilon})>_+^2 + \mu <\bm{\varepsilon}>_+:<\bm{\varepsilon}>_+ \right) + \left( \frac{\lambda}{2}  <tr(\bm{\varepsilon})>_-^2 + \mu <\bm{\varepsilon}>_-:<\bm{\varepsilon}>_- \right) \label{eq:eq7} \\
    \phi &=  Y_c h'(d) \dot{d} \label{eq:eq8}
\end{align}
where $\lambda$ and $\mu$ are the Lame's constants. Operators $<.>_+$ and $<.>_-$ indicate the positive and negative parts of the split and the expression for them is provided in Appendix \ref{ap:A}.  $g(d)$ is the energy degradation function such that $g(0) = 1, \;g(1), = 0$ and $g(d)$ convex on $[0, 1]$.  $Y_c$ is the crtitcal energy $[J/m^3]$. $h(d)$ is the damage softening function such that $h(0) = 0,\; h'(0)= 2,$ and $h(d)$ convex on $[0,1]$. Following \cite{Chevaugeon}, we use the below choice of $g(d)$ and $h(d)$.
\begin{align}
    g(d) &= (1-d)^2 + 0.1(1-d)d^3 \label{eq:eq9}\\
    h(d) &= 2d+3d^2 \label{eq:eq10}
\end{align}
The thermodynamic forces $(\bm{\sigma},Y)$ conjugate to the internal variables $(\bm{\varepsilon},d)$ are defined as follows:
\begin{align}
    \bm{\sigma} &= \frac{\partial \psi}{\partial \bm{\varepsilon}}  \label{eq:eq11}\\
    Y &= -\frac{\partial \psi}{\partial d} \label{eq:eq12}
\end{align}
where $Y$ is the local energy release rate.

The time domain of study $[0,T]$ is discretized into equal time intervals $t_0, t_1,...,t_n, t_{n+1},..T$ where $t_{n+1} = t_n + \Delta t$. The problem then involves finding the displacement field $\mathbf{u_{n+1}}$ and damage field $d_{n+1}$ at time step $t_{n+1}$, assuming $(\mathbf{u_n},d_n)$ are known from the previous time step. For the sake of simplicity, we drop the $n+1$ indices going forward. We employ the Newmark time integration scheme \cite{Newmark} to update the position and velocity at time $t_{n+1}$ as follows:
\begin{align}
    \mathbf{u} &= \mathbf{u_p} + \beta \Delta t^2 \mathbf{\ddot{u}} \label{eq:eq13}\\
    \mathbf{\dot{u}} &= \mathbf{\dot{u}_p} + \gamma \Delta t \mathbf{\ddot{u}} \label{eq:eq14}
\end{align}
where $\beta$ and $\gamma$ are the Newmark parameters  that determine the nature of numerical integration. The predictors are given as follows
\begin{align}
    \mathbf{u_p} &= \mathbf{u_n} + \Delta t \mathbf{\dot{u}_n} + \frac{\Delta t^2}{2} (1-2 \beta) \mathbf{\ddot{u}_n} \label{eq:eq15}\\
    \mathbf{\dot{u}_p} &= \mathbf{\dot{u}_n} + \Delta t (1-\gamma) \mathbf{\ddot{u}_n} \label{eq:eq16}
\end{align}

The mechanical problem to find $(\mathbf{u},d)$ at time $t_{n+1}$ can now be formulated as the following optimization problem:
\begin{align}
    \mathbf{u},d = \arg  \underset{d' \in \mathcal{A}_n}{\underset{\mathbf{u}' \in \mathcal{U}_n  }{  \min}} F(\mathbf{u}',d')  \label{eq:eq17}
\end{align}
with the incremental potential $F$ given by
\begin{align}
    F = \int_{\Omega} \left[ \frac{1}{2} \rho \frac{(\mathbf{u}-\mathbf{u_p})^2}{\beta \Delta t^2} + \psi (\bm{\varepsilon}(\mathbf{u}), d) + Y_c h(d) \right] d\Omega - \int_{\Gamma_N} \mathbf{t_N}.\mathbf{u} dS \label{eq:eq18}
\end{align}
The space $\mathcal{U}_n$ and $\mathcal{A}_n$ are defined as follows
\begin{align}
    \mathcal{U}_n &= \mathcal{U}(t_{n+1}) \label{eq:eq19}\\
    \mathcal{A}_n &= \{ d \in L^{\infty}(\Omega): d_n \leq d \leq 1 \}  \label{eq:eq20}
\end{align}
where $\mathcal{U}(t_{n+1})$ is given by Eq. (\ref{eq:eq2}).

It is now possible to calculate the directional derivative of the incremental potential $F$ with respect to the variations $(\mathbf{u^*},d^*)$ as follows:
\begin{align}
    DF(\mathbf{u},d) =  \int_{\Omega}  \rho \frac{(\mathbf{u}-\mathbf{u_p})}{\beta \Delta t^2}  \mathbf{u^*}  + \bm{\sigma} :  \bm{\varepsilon}(\mathbf{u^*}) -\left [Y +Y_c h'(d) \right] d^*d\Omega\;-\int_{\Gamma_N} \mathbf{t_N} . \mathbf{u^*} dS   \label{eq:eq21}
\end{align}
where the relation between thermodynamic forces and internal variables is used (see Eq. (\ref{eq:eq11},\ref{eq:eq12})). The admissible space for the variation in damage $d^*$ is given as follows:
\begin{align}
    \mathcal{A}^* = \{\delta d \in L^{\infty}(\Omega): \delta d \geq 0\} \label{eq:eq22}
\end{align}

The stationarity condition for the minimization problem given by Eq. (\ref{eq:eq17}) then reads
\begin{align}
    DF(\mathbf{u},d) \geq 0 \label{eq:eq23}
\end{align}
Zeroing the directional derivative with respect to $\mathbf{u^*}$ yields the following
\begin{align}
    \int_{\Omega}  \rho \frac{(\mathbf{u}-\mathbf{u_p})}{\beta \Delta t^2}  \mathbf{u^*}  + \bm{\sigma} :  \bm{\varepsilon}(\mathbf{u^*})d\Omega\;=\int_{\Gamma_N} \mathbf{t_N} . \mathbf{u^*} dS  \label{eq:eq24}
\end{align}
Notice that Eq. (\ref{eq:eq24}) is the time-discretized version of the equilibrium equation Eq. (\ref{eq:eq3}), where the expression for acceleration $\ddot{\mathbf{u}}$ is given by Eq. (\ref{eq:eq13}).  Now, the back substitution of the equilibrium equation (Eq. (\ref{eq:eq24})) in the stationarity condition (Eq. (\ref{eq:eq23})) gives the following
\begin{align}
    -\left[Y+Y_ch'(d)\right]d^* \geq 0 \implies Y-Y_ch'(d) \leq 0 \;\; \forall \;\; d^*\in \mathcal{A}^* \label{eq:eq25}
\end{align}
The evolution laws for damage can then be written using the following Karush-Kahn-Tucker conditions:
\begin{align}
    \dot{d} \geq 0,  \;\; Y-Y_ch'(d) \leq 0,\;\; (Y-Y_ch'(d))\dot{d} =0 \label{eq:eq26}
\end{align}
Therefore, the optimization problem given by Eq. (\ref{eq:eq17}) results in the solution satisfying the equilibrium and damage evolution equations.

\subsection{Lip-field regularization}
\label{sec:lf}

The problem of solving Eq. (\ref{eq:eq17}) within the space $\mathcal{A}_n$ is an ill-posed problem. Numerically, this results in a solution exhibiting undesirable sensitivity to mesh resolution, when attempted to solve using finite element techniques. This can be overcome by regularization techniques that introduce length scale(s) or time scales into the model. In the present paper, the focus is on using the Lip-field approach to introduce a length scale $l$ into the model. In particular, the Lip-field approach uses a Lipschitz regularization space $\mathcal{L}$ and constrains the damage field to lie in this space. The Lipschitz regularization space $\mathcal{L}$ is defined as follows:
\begin{align}
    \mathcal{L} = \left \{ d \in L^{\infty}(\Omega) : | d(\mathbf{x}) - d(\mathbf{y}) | \leq \frac{1}{l} dist(\mathbf{x},\mathbf{y}) \;\; \forall  \mathbf{x},\mathbf{y} \in \Omega  \right \}  \label{eq:eq27}
\end{align}
The regularized problem can then be rewritten as 
\begin{align}
    \mathbf{u},d = \arg  \underset{d' \in \mathcal{A}_n \cap \mathcal{L}}{\underset{\mathbf{u}' \in \mathcal{U}_n  }{  \min}} F(\mathbf{u}',d') \label{eq:eq28}
\end{align}
Notice that, along with the damage irreversibility constraint, the damage gradient is now bounded by enforcing $d$ to lie in $\mathcal{A}_n \cap \mathcal{L}$.

It is to be noted that $F$ is not convex with respect to $(\mathbf{u},d)$ although the admissible spaces for displacement and damage fields are convex. Hence, several local minima are possible. However, $F$ is convex with respect to $\mathbf{u}$ for a fixed $d$, and convex with respect to $d$ for a fixed $\mathbf{u}$ (provided $g(d)$ and $h(d)$ are convex).  Hence, a staggered minimization process is carried out as follows:
\begin{align}
    \mathbf{u^{k+1}} &=   \arg  {\underset{\mathbf{u}' \in \mathcal{U}_n  }{  \min}} F(\mathbf{u}',d^k) \label{eq:eq29} \\
    d^{k+1} &= \arg  {\underset{d' \in \mathcal{A}_n \cap \mathcal{L}}{  \min}} F(\mathbf{u^{k+1}},d') \label{eq:eq30}
\end{align}
In the case of the implicit Newmark scheme, the above iteration has to be repeated until convergence for a given time step. As will be seen shortly, for an explicit Newmark scheme, only one such iteration would be sufficient for a given time step.

\subsection{Lip-damage solver}
\label{ls}
\label{sec:bounds}
In this section, the focus is on the optimization process to find the damage variable. Finding $d$ can be made relatively more efficient by performing the minimization given by Eq. (\ref{eq:eq30}) only over a small portion of the domain $\Omega$, thanks to the bounds estimate proved in \cite{Moes LF}. The steps involved in finding $d$ in an efficient way are as follows:

\begin{enumerate}
    \item First, Eq. (\ref{eq:eq30}) is solved without the Lipschitz constraint. The problem then becomes local as no length scale is present. It can be written as
    \begin{align}
        d_{loc} =  arg \min_{d' \in A_n} F(\mathbf{u},  d') 
        \label{eq:eq31}
    \end{align}
    \item The predicted local damage field is projected into upper and lower bounds \cite{Moes LF, Chevaugeon} as follows:
    \begin{equation}
        \begin{aligned}
        \Bar{d}(\mathbf{x}) &= \max_{\mathbf{y} \in \Omega} ( d_{loc} (\mathbf{y}) -\frac{1\;}{l} dist(\mathbf{x},\mathbf{y})) \hspace{.8cm} &\forall \mathbf{x} \in \Omega\\
    \underline{d}(\mathbf{x})  &= \min_{\mathbf{y} \in \Omega} ( d_{loc} (\mathbf{y}) +\frac{1\;}{l} dist(\mathbf{x},\mathbf{y})) \hspace{.8cm} &\forall \mathbf{x} \in \Omega
    \end{aligned}
    \label{eq:eq32}
    \end{equation}
    where $dist(\mathbf{x},\mathbf{y})$ is the minimum distance between positions $\mathbf{x}$ and $\mathbf{y}$ inside $\Omega$. The upper and lower bounds have been proven to satisfy the following properties 
    \begin{align}
        &d_n(\mathbf{x}) \leq \underline{d}(\mathbf{x}) \leq d_{loc}(\mathbf{x}) \leq \Bar{d}(\mathbf{x}) \leq 1  \label{eq:eq33}\\ &\underline{d}(\mathbf{x})\leq d(\mathbf{x}) \leq \Bar{d}(\mathbf{x}) \label{eq:eq34}\\ &\overline{d}(\mathbf{x})=\underline{d}(\mathbf{x})  \implies d_{loc}(\mathbf{x})=d(\mathbf{x}) \in \mathcal{L} \label{eq:eq35}
    \end{align}
    Notice from Eq. (\ref{eq:eq35}) that in the regions where the bounds are equal, the local damage field $d_{loc}$ is the same as the solution of Lipschitz regularized damage field $d$.
    \item The problem now remains to find damage where the bounds are not equal. Now Eq. (\ref{eq:eq30}) is solved only in this small portion of the domain where the bounds are not equal.
\end{enumerate}
The local damage solver to find $d_{loc}$ given by Eq. (\ref{eq:eq31}) is quite fast as the minimization process is purely local. In contrast, due to the non-locality introduced by Lipschitz space, the minimization to find $d$ given by Eq. (\ref{eq:eq30}) is considered relatively more expensive in terms of computational effort. However, the use of the bounds greatly increases the efficiency of the Lip-field damage solver by performing the non-local minimization process only over a small portion of the domain $\Omega$, where the bounds are not equal. This is one of the originalities of the Lip-field approach.

\section{Numerical resolution}
\label{nr}
We limit our discussion here to Newmark's central difference scheme (explicit) as it has been widely accepted for capturing material behavior under dynamic loading conditions. For this case, we use $\beta =0$ and $\gamma = .5$. Hence, Eq.(\ref{eq:eq13}) and Eq. (\ref{eq:eq14}) results in the following:
\begin{align}
    \mathbf{u} &= \mathbf{u_p}  \label{eq:eq36}\\
    \mathbf{\dot{u}} &= \mathbf{\dot{u}_p} + \frac{\Delta t}{2} \mathbf{\ddot{u}} \label{eq:eq37}
\end{align}
with the following expression for the predictors (as per Eq. (\ref{eq:eq15}) and Eq. (\ref{eq:eq16}))
\begin{align}
    \mathbf{u_p} &= \mathbf{u_n} + \Delta t \mathbf{\dot{u}_n} + \frac{\Delta t^2}{2}  \mathbf{\ddot{u}_n} \label{eq:eq38} \\
    \mathbf{\dot{u}_p} &= \mathbf{\dot{u}_n} + \frac{\Delta t}{2} \mathbf{\ddot{u}_n} \label{eq:eq39}
\end{align}

\subsection{Space discretisation}
Consider the finite element space discretization $\Omega^h$ of the domain $\Omega$ using linear triangular elements. This allows approximating the kinematic fields using piecewise linear basis functions as follows:
\begin{align}
    \mathbf{u^h}(\mathbf{x}) &= \sum_{i \in S} \mathbf{u_i}(t) \mathbf{N_i}(x) \label{eq:eq40} \\
    \mathbf{\dot{u}^h}(\mathbf{x}) &= \sum_{i \in S} \mathbf{\dot{u}_i}(t) \mathbf{N_i}(x) \label{eq:eq41}\\
    \mathbf{\ddot{u}^h}(\mathbf{x}) &= \sum_{i \in S} \mathbf{\ddot{u}_i}(t) \mathbf{N_i}(x)  \label{eq:eq42}
\end{align}
where $S$ is the set of nodes on $\Omega^h$. Applying the space discretization to the equilibrium equation Eq. (\ref{eq:eq3}) results in the following
\begin{align}
    \mathbf{M}  \mathbf{\ddot{u}^h} = - \mathbf{F} + \mathbf{R} 
    \label{eq:eq43}
\end{align}
The corresponding mass matrices and force vectors are given below
\begin{align}
    \mathbf{M}_{ij} &= \int_{\Omega} \rho \mathbf{N}_i \mathbf{N}_j \;d\Omega  \label{eq:eq44}\\
    \mathbf{F}_i &= \int_{\Omega} \bm{\sigma}(\bm{\varepsilon},d) : \nabla_s \mathbf{N}_i \;d\Omega \label{eq:eq45}\\
    \mathbf{R_i} &= \int_{\Gamma_N} \mathbf{t_N} . \mathbf{N}_i \;dS \label{eq:eq46}
\end{align}
In this study, we preserve the mass matrix given by Eq. (\ref{eq:eq44}). In other words, a consistent mass matrix has been used (the mass matrix is not lumped).
For the minimization process to find $d$ (given by Eq. (\ref{eq:eq30})) in discrete form, a lip-mesh $\Delta^h$ is constructed (see \cite{Chevaugeon}) by joining the element centroids of the original mesh $\Omega^h$. The damage field is then stored as a vector $\mathbf{d^h}$, where the elements of the vector represent the damage values at the lip-mesh vertices (or at element centroids of the original mesh). The lip-mesh allows approximating the continuous spaces $\mathcal{A}_n$ and $\mathcal{L}$ by discrete spaces over $\Delta^h$. The use of the lip-mesh also ensures that the number of Lipschitz constraints is kept minimum. Note that, the lip-mesh has to be constructed only once at the beginning of a simulation for a given mesh $\Omega^h$.  In order to estimate the bounds, a Dijkstra-based fast marching algorithm, as described in \cite{Chevaugeon}, is utilized. The local minimization process to find $d_{loc}$ is executed utilizing Python's \textit{scipy} package \cite{scipy}, while the non-local minimization to determine the Lipschitz-regularized damage field is accomplished using the \textit{cvxopt} package of Python \cite{cvxopt}. We have limited our discussion on lip-damage solver in continuous setting (Section \ref{sec:bounds}) and one may refer to \cite{Chevaugeon, Rajasekar} for a detailed discussion on the implementation of lip-damage solver in discrete setting.

The algorithm to find the displacement field and damage field is provided in Algorithm \ref{alg:cap}

\begin{algorithm}[H]
\caption{Explicit scheme}\label{alg:cap}
\begin{algorithmic}
\State Initialize $t \gets 0$, $\mathbf{u^h_0}, \mathbf{\dot{u}^h_0}, \mathbf{\ddot{u}^h_0}, \mathbf{d^h_0}, \Delta t$
\State Assemble the consistent mass matrix $\mathbf{M}$ \Comment{using Eq. (\ref{eq:eq44})}
\While{$t \leq T$}
    \State Find the predicted nodal values of displacement $\mathbf{{u^h_p}}$ and velocity $\mathbf{\dot{u}^h_p}$ \Comment{using Eq. (\ref{eq:eq38}, \ref{eq:eq39})}
    \State Find  nodal dispalcement vector $\mathbf{u^h}$  \Comment{usign Eq. (\ref{eq:eq36})}
    \State Determine local damage field vector $\mathbf{d^h_{loc}}$ \Comment{usign Eq. (\ref{eq:eq31})}
    \State Compute the upper and lower bound vectors ${{\mathbf{\Bar{d}^h}}},\mathbf{\underline{d}^h}$ \Comment{usign Eq. (\ref{eq:eq32})}
    \State set $\mathbf{d^h} = \mathbf{d^h_{loc}}$ in the regions where $\mathbf{\Bar{d}^h}=\mathbf{\underline{d}^h}$ \Comment{using Eq. (\ref{eq:eq35})}
    \State Find $\mathbf{d^h}$ in the regions where $\mathbf{\Bar{d}^h} \neq \mathbf{\underline{d}^h}$ using Eq. (\ref{eq:eq30})
    \State Compute the force vectors $\mathbf{F}$ and $\mathbf{R}$ \Comment{using Eq. (\ref{eq:eq45}, \ref{eq:eq46})}
    \State Compute the nodal acceleration $\mathbf{\ddot{u}^h}$ using equilibrium equation \Comment{using Eq. (\ref{eq:eq43})}
    \State Compute nodal velocity vector $\mathbf{\dot{u}^h}$ \Comment{usign Eq. (\ref{eq:eq37})}
    \State $t \gets t+ \Delta t$
\EndWhile
\end{algorithmic}
\end{algorithm}

\section{Simulation results}
\label{sr}
This section presents the numerical results for the two-dimensional cases, which evaluate the capability of the Lip-field approach to simulate dynamic brittle fracture. The plane strain assumption is employed in all the tests conducted. This assumption leads to the following equations to determine the Lame's constants ($\lambda,\mu$): $\lambda = E \nu /(1+\nu)/(1-2 \nu)$ and $\mu = E/2/(1+\nu)$, where $E$ is Young's modulus and $\nu$ is the Poisson's ratio. The meshes used for the simulation are created using the open source \textit{gmsh} software \cite{gmsh}, while the associated lip-mesh is generated using Python's \textit{triangle} package \cite{triangle}. The code required to obtain the simulation results can be downloaded from \textit{https://github.com/GRajasekar28/lip-field-dynamics}. The expression for the dilatational wave speed $c_d$, shear wave speed $c_s$, and Rayleigh wave speed $c_R$ are given as follows:
\begin{align}
    c_d = \sqrt{\frac{\lambda+2\mu}{\rho}}, \;\;\;\;\;\;\; c_s = \sqrt{\frac{\mu}{\rho}} ,\;\;\;\;\;\;\; c_R = \frac{.862+1.14\nu}{1+\nu} c_s\label{eq:eq47}
\end{align}
As per the fracture mechanics theory, the limiting crack speed is the Rayleigh wave
speed \cite{Freund}.  Moreover, the considered explicit scheme is only conditionally stable. Hence, the time step $\Delta t$ is determined by the Courant-Freidrichs-Lewy (CFL) condition, such that $\Delta t < \Delta t_c = h/c$, where $\Delta t_c$ is the critical time step, $h$ is the size of the smallest element in the mesh and $c$ is the material sound speed. Moreover, for a given mesh, $\Delta t_c$ for a sound material is always smaller than a corresponding damaged material, due to the relatively higher speed of sound in the sound material. As a result, the time step $\Delta t$ is not affected by damaged elements.

\subsection{Single edge notched tension test}
\begin{figure}[H]
  \centering
  {\includegraphics[trim=1cm 19cm 1cm 1cm,clip,scale= .8]{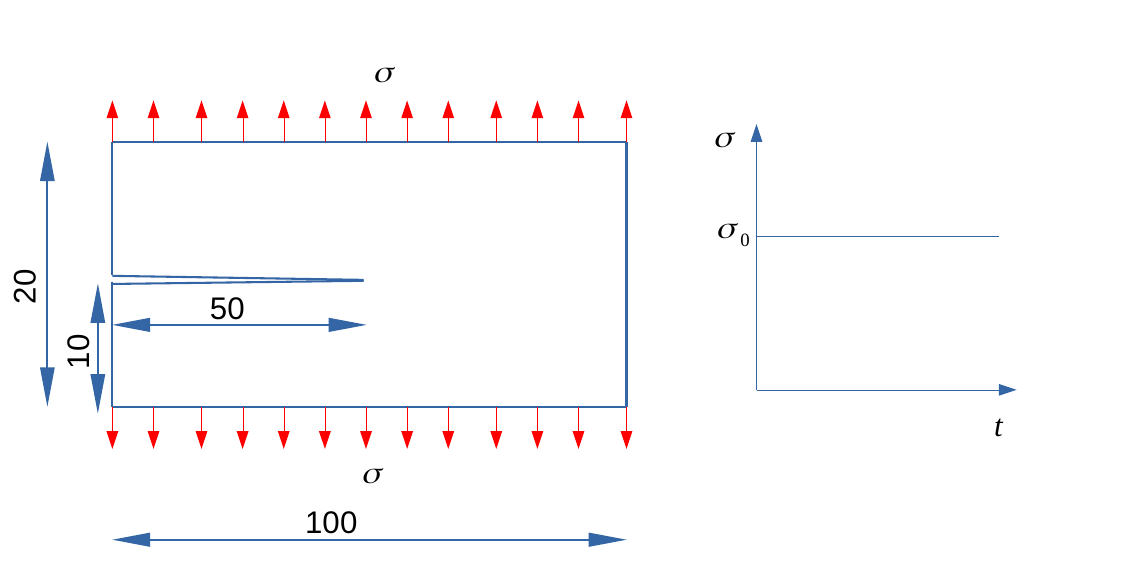}}
  \caption{Geometry of single edge notched tension test (left) and the loading history (right). Dimensions of the geometry are in millimeter ('$mm$')} \label{fig:1}
\end{figure}

In this example, we consider a rectangular block that has a pre-existing crack, which is subjected to dynamic tensile loading as shown in Figure \ref{fig:1}. This problem has been commonly considered in the fracture mechanics community to study the dynamic crack branching phenomenon. Experimental results on similar tests were reported  in \cite{Ramulu, Sharon}. Several authors in the literature have attempted to address this problem numerically. For instance, in \cite{Song},  CZM is used to model the dynamic branching for the specimen under consideration. In \cite{Moreau}, the authors utilized the TLS approach to simulate crack branching in a similar test, whereas the works reported in \cite{Borden, Marigo, Wu} present outcomes for the phase-field model applied to the same test.
The material properties used for the simulation are as follows: $E$=32 GPa, $\nu$=.2, $\rho$ = 2450 $kg/m^3$ and $G_c$ =3 N/m, where $G_c$ is the Griffith fracture energy release rate. The material properties considered result in the following values for the wave speeds: $c_d$=3810 m/s, $c_s$ = 2332 m/s, and $c_R$=2119 m/s.  To accelerate the computation process, we make use of the symmetry and consider only the bottom half of the problem, and set $\mathbf{u}.\mathbf{e_2} = 0$ on the top surface excluding crack.

For the present case, $\bm{\sigma}_0$ = 1 MPa is used.
For the considered degradation function $g(d)$ and damage softening function $h(d)$, the relation between $G_c$ and $Y_c$ is given as follows \cite{Chevaugeon}:
\begin{align}
    G_c = 4Y_cl \label{eq:eq48}
\end{align}
The Lipschitz regularizing length scale is set $l$ = 1.25 mm for this case. The relation between $G_c$ and $Y_c$ given by Eq. (\ref{eq:eq48})  results in $Y_c$ = 600 $J/m^3$.
 Three different meshes are considered for the study as shown in Table \ref{Table:1}. The meshes have been refined in the right half of the domain (with the smallest element size being $h$), where the crack is expected to propagate. The time step $\Delta t$ for each mesh is chosen such that $\Delta t$ = .8 h/c.
 \begin{table}[H]
\begin{center}
\caption{Mesh sizes and characteristic length scale used for the study}
\label{Table:1}       
\begin{tabular}{llll}
\hline\noalign{\smallskip}
Mesh  & h (mm) & l (mm) & l/h\\
\noalign{\smallskip}\hline\noalign{\smallskip}
Mesh 1 & .5 & 1.25 & 2.5 \\
\noalign{\smallskip}\hline\noalign{\smallskip}
 Mesh 2 & .25 & 1.25 & 5 \\
 \noalign{\smallskip}\hline\noalign{\smallskip}
 Mesh 3 & .166667 & 1.25 & 7.5 \\
 \noalign{\smallskip}\hline
\end{tabular}
 \end{center}
\end{table}

\begin{figure}[H]
  \centering
  {\includegraphics[trim=1cm 6cm 1cm 1cm,clip,scale= .6]{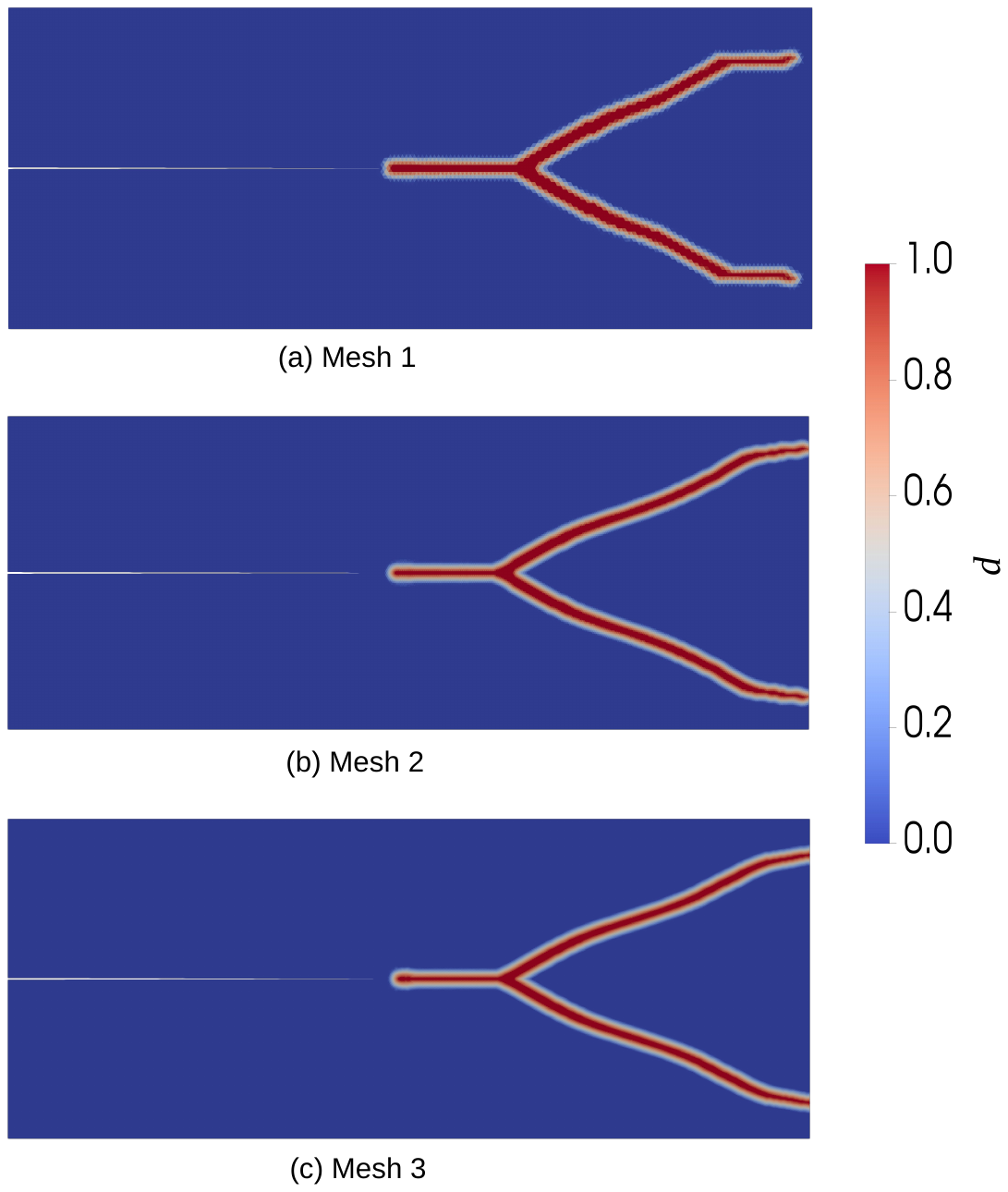}}
  \caption{Damage field for three different meshes at $t=80\mu s$ } \label{fig:2}
\end{figure}

 Figure \ref{fig:2} shows the damage field for three different meshes at $t=80\mu s$. With increasing time, the crack starts to propagate from the notch toward the right. Eventually, the crack branches into two separate cracks as observed in experiments. Moreover, the crack path appears to be consistent across all the meshes. However, the crack fails to reach the boundary for mesh 1 in the same amount of time. This can be attributed to the coarse mesh, which is not adequate enough to resolve the length scale. It is to be noted that the simulation results effectively replicate the key characteristics of the crack path observed in the experiment \cite{Ramulu}.
 
\begin{figure}[H]
  \centering
  {\includegraphics[trim=1cm 19cm 1cm 3.5cm,clip,scale= .8]{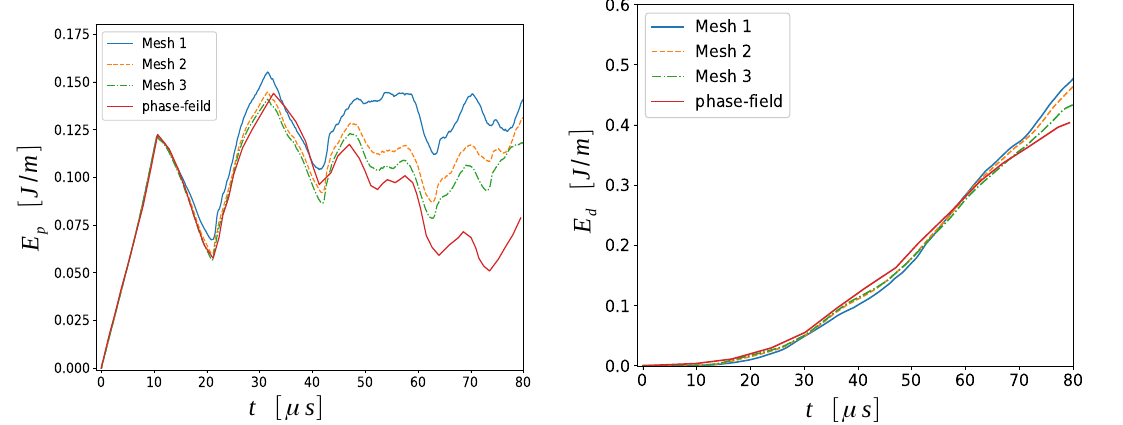}}
  \caption{Plots of potential energies over time (left) and dissipated energies over time(right). Lip-field results (for different meshes) compared with phase-field results \cite{Borden}} \label{fig:3}
\end{figure}

\begin{figure}[H]
  \centering
  {\includegraphics[trim=1cm 8cm 1cm 13cm,clip,scale= .8]{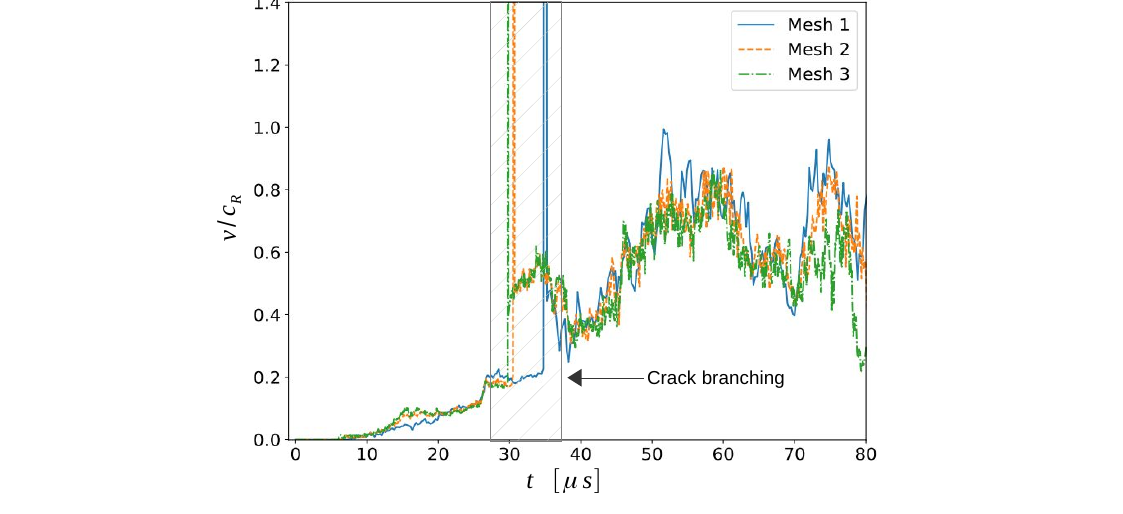}}
  \caption{Plot of crack tip velocity over time} \label{fig:4}
\end{figure}

Figure \ref{fig:3} plots the global values of potential energy $E_p$ and dissipated energies $E_d$ given as follows:
\begin{align}
    E_p &= \int_{\Omega} \psi(\bm{\varepsilon},d) \;d\Omega \label{eq:eq49}\\
    E_d &= \int_{\Omega} Y_c h(d)\; d\Omega \label{eq:eq50}
\end{align}
It can be noticed that as the mesh is refined, the plots for the energies start to converge. Additionally, as depicted in the figure, the profiles of the potential and dissipated energies exhibit a significant resemblance to the findings of the phase-field model \cite{Borden} for a majority of the time duration. 
The ratio of crack-tip velocity to Rayleigh wave speed as a function of time is plotted in Figure \ref{fig:4}. Crack tip velocity is evaluated as a rate of change of crack length.  As the crack tip is not tracked algorithmically, the crack length is measured with a post-processing step. For a material with a single crack, under the assumption that the damage field saturates the Lispchitz constraint, the crack length is given by $\int_{\Omega} d/l\; d\Omega$. However, in the case being considered where there is symmetric branching (as only one-half of the problem is solved), the crack length at time $t$ is evaluated as $a = a_n + \Delta a$, where $\Delta a$ is the increase in crack length in time $\Delta t$. $\Delta a$  is given as follows
\begin{equation}
\begin{aligned}
    \Delta a = \int_{\Omega} \frac{\Delta d}{l} \; d\Omega \;\;\; if \;\;\; t<t_{br}\\
    \Delta a = \int_{\Omega} \frac{\Delta d}{2l} \; d\Omega \;\;\;  if \;\;\; t\geq t_{br}
\end{aligned}
\label{eq:eq51}
\end{equation}
where $\Delta d = d - d_n$ and  $t_{br}$ is the time step associated with the onset of branching and is found using a post-processing step.

\begin{figure}[H]
  \centering
  {\includegraphics[trim=1cm 20cm 1cm 3cm,clip,scale= .8]{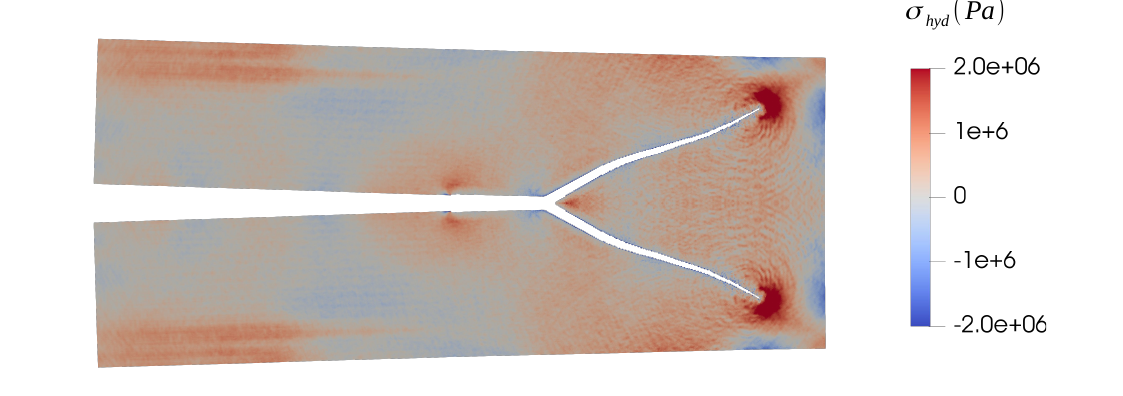}}
  \caption{Plot of hydrostatic stress over the deformed mesh at $t=70\;\mu s$ (magnification factor =50). To show the crack, elements with $d>.99$ has been removed.} \label{fig:5}
\end{figure}

As shown in Figure \ref{fig:4}, the crack tip velocity initially increases gradually and then abruptly goes beyond Rayleigh wave speed $c_R$ and then decreases again to propagate below $c_R$. From post-processing, crack branching was observed approximately at $t= (36,31,30.6)\;\mu s $ for three different meshes respectively. It is to be noted that, these values can be attributed to the time steps with peak crack tip velocity as indicated in Figure \ref{fig:4}. It can also be seen that the crack tip velocity is below Rayleigh wave speed for all the time steps excluding the time steps corresponding to crack branching. 
The sudden increase in crack velocity beyond the Rayleigh wave speed at the onset of crack branching could potentially be attributed to the damage widening. In this scenario, the estimated increase in crack length provided by Equation (\ref{eq:eq51}) might be overestimated and hence result in spurious values for crack tip velocity.  Figure \ref{fig:5} plots the hydrostatic stress for mesh 3 over the deformed configuration. 

\subsection{Influence of load intensity}
\begin{figure}[H]
  \centering
  {\includegraphics[trim=1cm 12cm 1cm 4cm,clip,scale= .7]{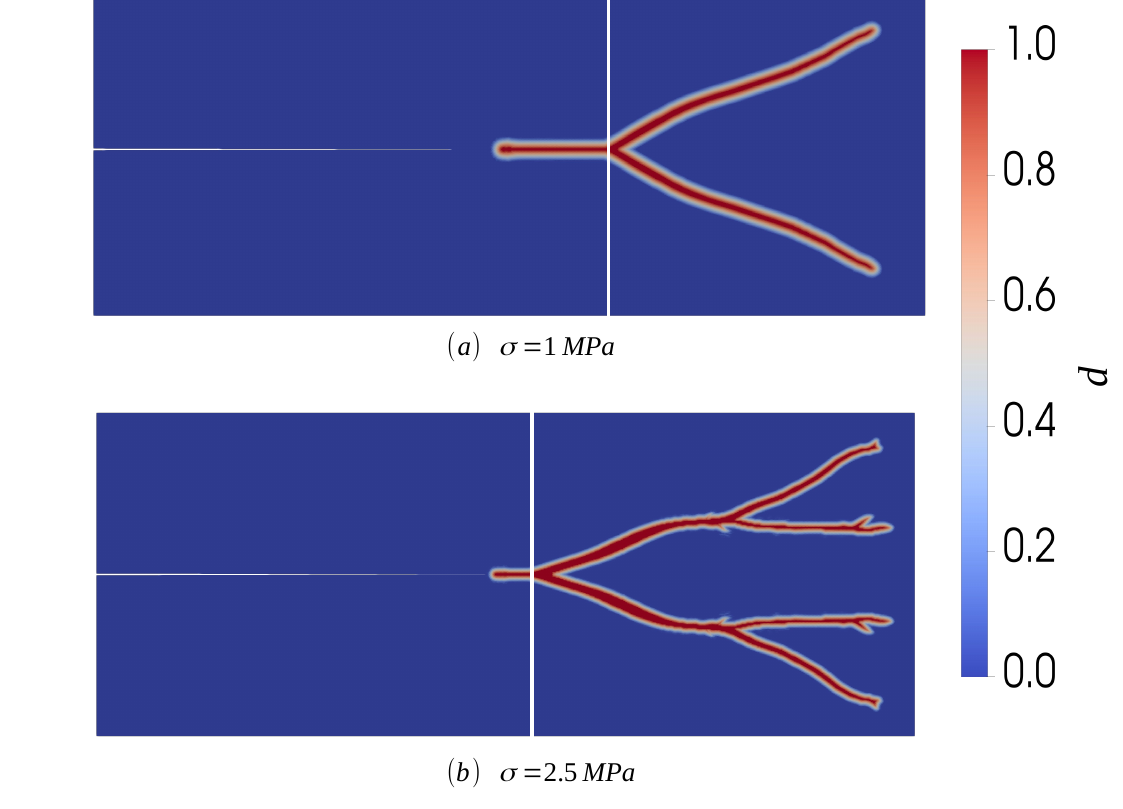}}
  \caption{Influence of load intensity on crack. The position where crack branching begins is indicated by white lines.} \label{fig:6}
\end{figure}

To evaluate the ability of the Lip-field approach in capturing multiple crack branches, we consider Mesh 3 with the Lipschitz regularizing length scale set as $l=.8333\;mm$. This leads to the following value for critical volumetric energy release rate: $Y_c = 899.999 \;J/m^3$. In this case, we apply two different traction forces $\bm{\sigma}_0 = 1\; MPa$ and $\bm{\sigma}_0 = 2.5\; MPa$. The corresponding results for the damage profile are compared in Figure \ref{fig:6}. It can be seen that crack branching occurs very close to the notch tip for the latter case, however, the crack branching is delayed for the former. Additionally, as seen in the phase-field models  \cite{Wu}, multiple crack branches are also produced when the applied traction force is increased. It is also evident from this simulation that the Lip-field approach is able to capture the multiple crack branches.

\section{Kalthoff-Winkler test}

\begin{figure}[H]
  \centering
  {\includegraphics[trim=1cm 5cm 1cm 13cm,clip,scale= .7]{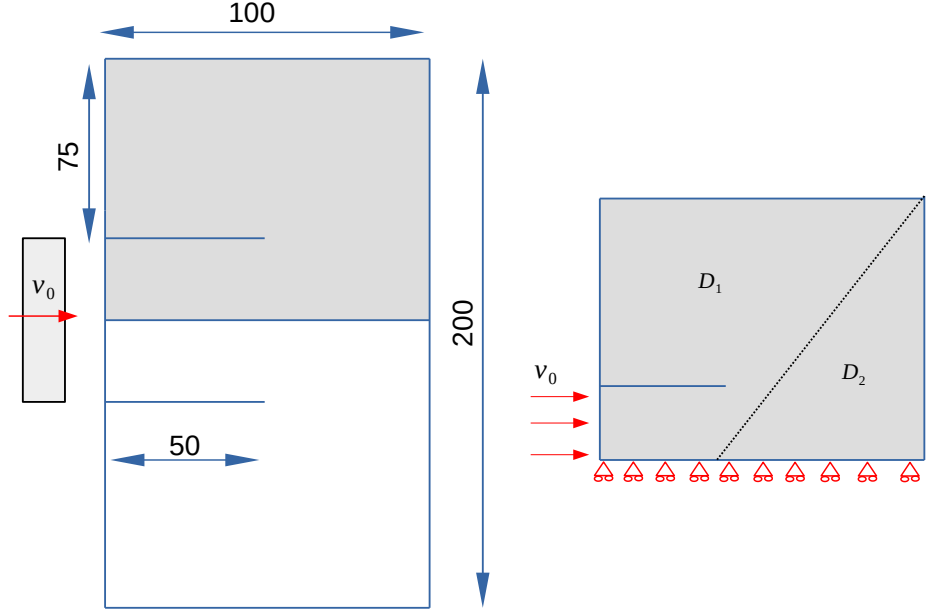}}
  \caption{Geometry and boundary conditions for the Kalthoff-Winkler experiment. Dimensions of the geometry are in millimeter ('$mm$')} \label{fig:7}
\end{figure}

In this section, we are interested in modeling crack initiation and propagation under dynamic shear loading. This experiment concerns a maraging steel 18Ni1900 plate with 2 pre-existing cracks impacted by a projectile as shown in Figure \ref{fig:7}.  The material properties used for the simulation are as follows: $E$  = 190 GPa, $\nu$ =.3, $\rho$ =8000 $kg/m^3$ and $G_c$ = 22.2E3 N/m. This lead to the following values for the wave speeds: $c_d$ = 5654 m/s, $c_s$ = 3022 m/s, and $c_R$ = 2799 m/s. 

In the experiment, two different failure modes were observed as impact velocity changes: (a) at low impact velocities, a brittle fracture was observed with a crack propagation angle of about $70^o$ (b) when the impact velocity increases, the ductile failure phenomenon was observed with a mode-II crack propagating at an angle of about $-10^o$. 
The Kalthoff-Winkler experiment is noteworthy because it demonstrates that as the impact velocity increases, the way in which a material fails changes from brittle to ductile failure. This finding goes against the conventional assumption that the failure mode would shift from ductile to brittle as the strain rate increases. 

Numerical results of a similar experiment have been reported in the literature using CZM \cite{Song}, TLS approach \cite{Moreau}, and phase-field approach \cite{Borden, Marigo, Wu}.

Due to the symmetry of the geometry and the loading conditions, only the upper half of the problem is being considered as shown in Figure \ref{fig:7}.
Our assumption is that the elastic impedance of the specimen and the projectile are identical. Therefore, we only need to apply half of the initial velocity of the projectile as the initial condition: $\dot{\mathbf{u}} . \mathbf{e}_1 = v0/2$. We consider two different impact velocities for the study (i) $v0$ = 33 $m/s$, (ii) $v0$ = 100 $m/s$.  A uniform mesh with triangular elements of size $h$ is used for both simulations as shown in Table \ref{Table:2}. The time step $\Delta t$ is chosen such that $\Delta t = .9 h/c$. The Lipschitz regularizing length scale is set as $l$=2 mm. Eq. (\ref{eq:eq48}) then yields $Y_c$ = 2775E3 $J/m^3$.
\begin{table}[H]
\begin{center}
\caption{Mesh sizes and characteristic length scale used for Kalthoff-Winkler experiment}
\label{Table:2}       
\begin{tabular}{llll}
\hline\noalign{\smallskip}
 Mesh  & h (mm) & l (mm) & l/h\\
\noalign{\smallskip}\hline\noalign{\smallskip}
Mesh 1 & .3333 & 2 & 6 \\
 \noalign{\smallskip}\hline
\end{tabular}
 \end{center}
\end{table}

\subsection{Case A}
\begin{figure}[H]
  \centering
  {\includegraphics[trim=1cm 12cm 1cm 3cm,clip,scale= .8]{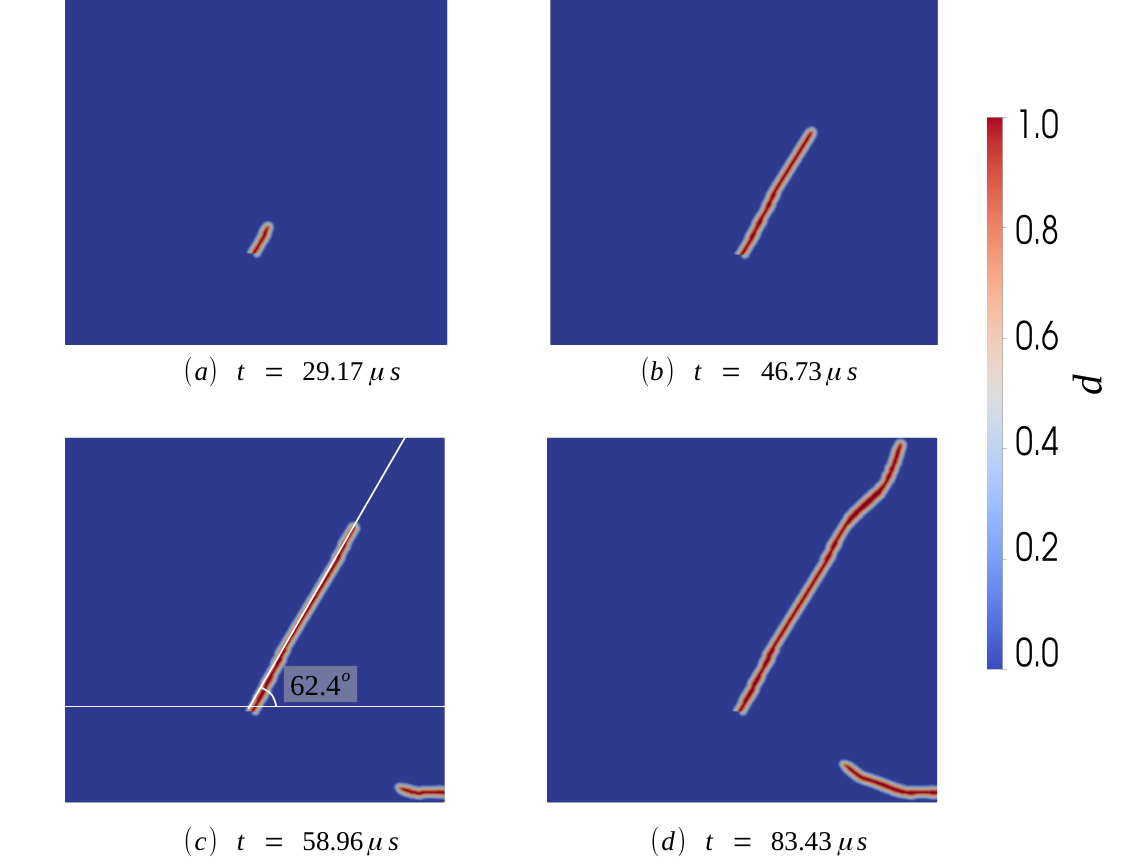}}
  \caption{Evolution of damage profiles for Kalthoff-Winkler experiment} (all dimensions in $mm$) \label{fig:8}
\end{figure}

Here, we consider the initially imposed velocity $v0 = 33\;m/s$. The evolution of the damage field is shown in Figure \ref{fig:8}. The impact due to the projectile initially creates a compressive wave in the plate, which creates a predominantly mode-II dynamic loading at notch tips. At the start, a mode-I crack begins at the tip of the notch and subsequently extends at an angle of roughly $62.4^o$ degrees to the notch line. This result is reasonably consistent with the experimental findings, which indicate a crack propagation angle of approximately $70^o$.

Moreover, the presence of secondary cracks at the bottom right edge is also observed. This could be due to the wave reflections occurring at the bottom right edge. Nevertheless, the Kalthoff-Winkler experiments do not mention the occurrence of such secondary cracks. In contrast, some numerical techniques that rely on damage mechanics-based approaches reveal the presence of such secondary cracks \cite{Moreau},\cite{Belytschko}.
\begin{figure}[H]
  \centering
  {\includegraphics[trim=1cm 3.3cm 1cm 13.5cm,clip,scale= .7]{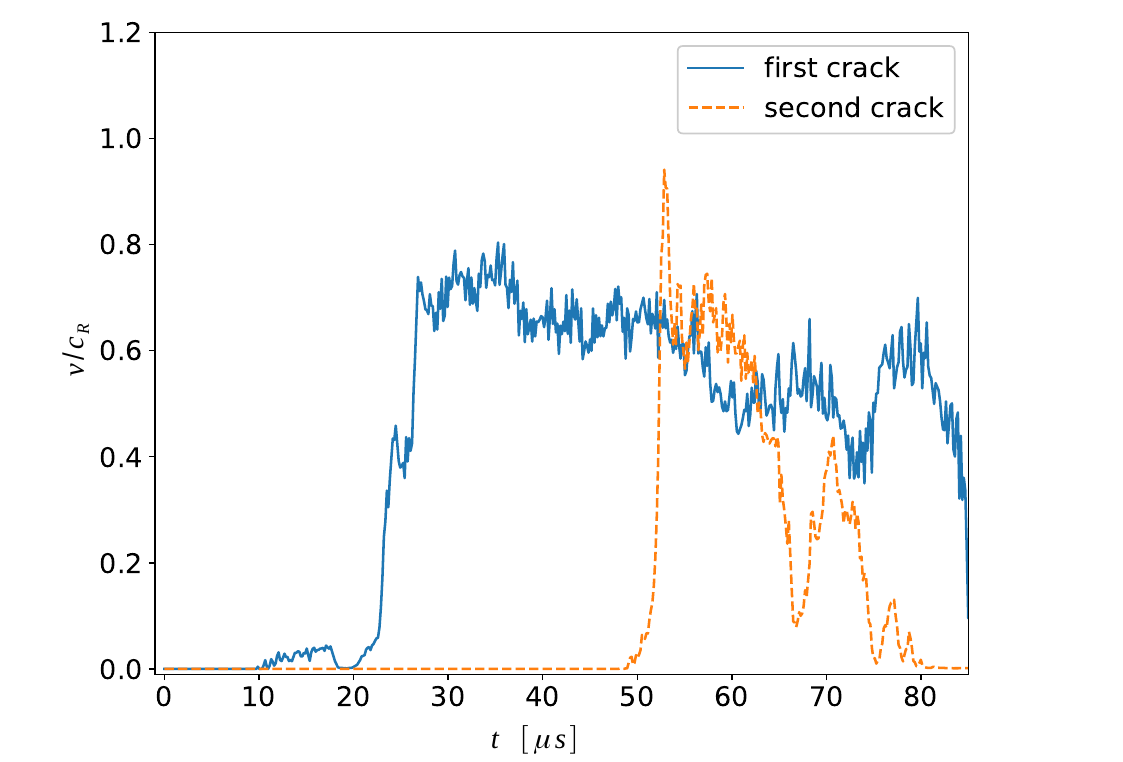}}
  \caption{Plot of crack tip velocity over time} \label{fig:9}
\end{figure}
The ratio of crack tip velocity to Rayleigh wave speed is plotted in Figure \ref{fig:9} for both the first and second crack. The crack velocity is calculated as the rate of change of crack length. The crack length for both the first and second crack ($a_1,a_2$) is evaluated as follows:
\begin{equation}
    \begin{aligned}
        a_i =  \int_{D_i} \frac{d}{l}  \; d\Omega \;\;\;\;\;\; \forall \;\; i \in \{1,2\}
    \end{aligned}
    \label{eq:eq52}
\end{equation}
where $D_1$ and $D_2$  indicate the region of the domain  $\Omega$ as indicated in Figure \ref{fig:7}.
It is clear from Figure \ref{fig:9} that the crack velocity lies below Rayleigh wave speed in this case for both cracks. 
Furthermore, it is evident that the emergence of the secondary crack is postponed. The duration required for the wave originating from the impacted boundary to reach the opposite end of the specimen twice is given by $3L/c_d = 53.06 \mu s$, where $L$ is the length of the specimen. Also, notice from Figure \ref{fig:9} that the second crack starts to grow approximately at $52 \mu s$. This confirms the likelihood that the second crack is a result of wave reflection.

\subsection{Case B}
\begin{figure}[H]
  \centering
  {\includegraphics[trim=1cm 19cm 1cm 1cm,clip,scale= .7]{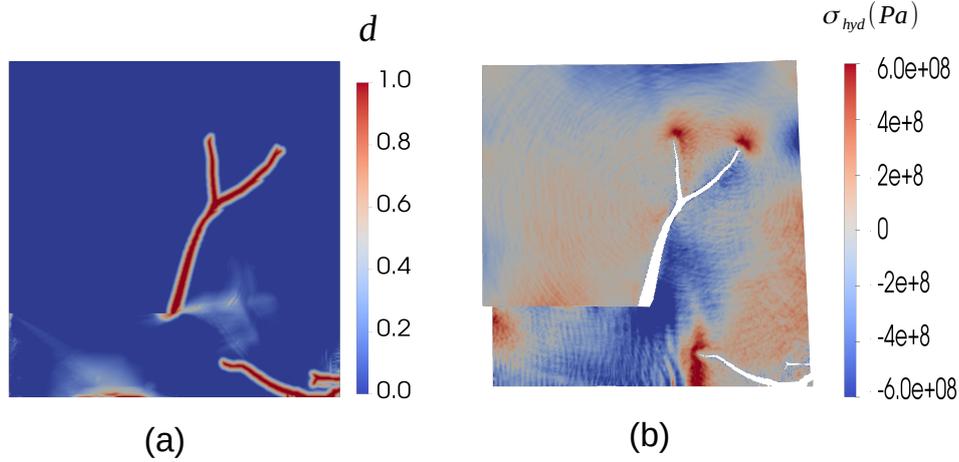}}
  \caption{Plots obtained for $t=4.8 \mu s$ (a) Damage profile and (b) hydrostatic stress over deformed configuration with magnification factor =5. To show crack, elements with $d>.99$ has been removed} \label{fig:10}
\end{figure}

In this study, we are interested in the impact velocity $v_0= 100\; m/s$. The respective damage field is plotted in Figure \ref{fig:10}a. As the impact velocity rises, successive crack branching and 
 the presence of multiple cracks at the bottom right corner are observed. Figure \ref{fig:10}b plots the hydrostatic stress on the deformed configuration. It is to be noted that a similar damage profile was also observed for the phase-field models \cite{Marigo}. However, in the case of phase-field models, multiple micro-crack branches (crack branches that don't grow) were observed along the primary crack. This is expected to be captured in the Lip-field model by refining the mesh for a fixed regularizing length scale $l$.
 As noted previously, the experimental findings show a ductile failure phenomenon for higher impact velocities, with shear fractures arising at an angle close to $-10^o$. However, our simulation results do not exhibit such behavior. The reason for this disparity is that the elastic-damage model used in the simulation does not consider the ductile effects. An elastic-plastic damage model would be better suited to simulate high-impact velocities.

\section{Conclusion}
\label{conc}

This paper extends the Lip-field approach from the one-dimensional case of dynamic fracture \cite{Moes LF dynamics} to the case of two dimensions. Unlike the traditional Thick-Level-Set (TLS) approach \cite{Moes TLS}, the Lip-field approach preserves the local solution and non-locality is introduced only when the Lipschitz constraints are not satisfied.

In the Lip-field approach, only the local equations are considered. These local equations are given by the minimization of local incremental potential. The main difference between the Lip-field approach with the phase-field approach lies in the way the local equations are regularized. In the Lip-field approach, the regularization is through Lipschitz constraints that the damage gradient is bounded by a critical value. However, in the phase-field approach, the regularization is by the addition of the damage gradient term to the incremental potential. The Lipschitz constraint introduces a length scale into the model. This helps in preserving the mathematical relevance of the problem.

We introduced an explicit time discretization method that is based on Newmark's central difference time integration scheme for the momentum equation. Additionally, we proposed the use of bounds, which were established in a previous study \cite{Moes LF}, to simplify the process of determining the damage field. By doing so, the local damage field can be maintained in the area of the domain where the bounds are the same, while the damage field only needs to be computed for the remaining portions of the domain, thereby significantly reducing the computational effort. Furthermore, in parallel computing, this method enables us to calculate the Lipschtiz regularized damage field separately in multiple regions where the bounds are not equal. This feature can be especially advantageous in fragmentation analyses, where the occurrence of multiple cracks is anticipated.

Finally, a variety of examples have been treated showing the ability of the Lip-field approach to model dynamic crack propagation behavior.

\section{Appendix} \label{ap:A}
In this section the expression for the operators $<.>_+$ and $<.>_-$ are provided. Using the eigen split of a tensor, the strain $\bm{\varepsilon}$ can be written as follows
\begin{align}
    \bm{\varepsilon} =  <\bm{\varepsilon}>_+ \; + \;< \bm{\varepsilon}>_{-}  \label{eq:eq60}
\end{align}
where $< \bm{\varepsilon}>_+$ and $< \bm{\varepsilon}>_{-}$ indicate the tensile and compressive modes. These modes are defined based on the eigen values ($\{ \lambda^a \}_{a=1,2,..,\beta}$) and eigen vectors ($\{ \mathbf{n}^a \}_{a=1,2,..,\beta}$) of $ \bm{\varepsilon}$. $\beta$ takes the values 2 and 3 for two and three dimensions respectively. Subsequently, the equation for the tensile and compressive modes can be presented as:
\begin{align}
    < \bm{\varepsilon}>_+ = \sum_{a=1}^{\beta} <\lambda^a>_+ \mathbf{n}^a \otimes \mathbf{n}^a \hspace{1cm} < \bm{\varepsilon}>_{-} = \sum_{a=1}^{\beta} <\lambda^a>_{-} \mathbf{n}^a \otimes \mathbf{n}^a \label{eq:eq53}
\end{align}
along with the following definitions of the bracket operators::
\begin{align}
    <p>_+ = \frac{1}{2}( p + |p|) \hspace{1cm} <p>_{-} = \frac{1}{2}( p - |p|)  \label{eq:eq54}
\end{align}

%

\end{document}